\input{aipcheck}
\documentclass[final]{aipproc}

\layoutstyle{8x11single}

\begin{document}

\title{Public infrastructure for Monte Carlo simulation :\\
publicMC@BATAN}


\author{A. A. Waskita\thanks{adhyaksa@batan.go.id}}{
  address={Center for Development of Nuclear Informatics\footnote{http://www.batan.go.id/ppin/}, National Nuclear Energy Agency, Kompleks Puspiptek Serpong, Tangerang 15310, Indonesia}
}
\author{N. A. Prasetyo}{
  address={Center for Applied Aerospace Technology, National Space and Aviation Agency\footnote{http://www.lapan.go.id}, Jl. Raya LAPAN - Rumpin, Tromol Pos 7, Serpong, Tangerang 15310, Indonesia}
  ,altaddress={Faculty of Computing Science\footnote{http://www.cs.ui.ac.id}, University of Indonesia, Kampus UI Depok, Depok 16424, Indonesia}
}
\author{Z. Akbar}{
  address={Group for Theoretical and Computational Physics, Research Center for Physics\footnote{http://teori.fisika.lipi.go.id}, Indonesian Institute of Sciences, Kompleks Puspiptek Serpong, Tangerang 15310, Indonesia}
  ,altaddress={Group for Bioinformatics and Information Mining\footnote{http://www.inf.uni-konstanz.de/bioml/}, Department of Computer and Information Science, University of Konstanz, Box D188, D-78457 Konstanz, Germany}
}
\author{L.T. Handoko\thanks{laksana.tri.handoko@lipi.go.id, handoko@teori.fisika.lipi.go.id}}{
  address={Group for Theoretical and Computational Physics, Research Center for Physics\footnote{http://teori.fisika.lipi.go.id}, Indonesian Institute of Sciences, Kompleks Puspiptek Serpong, Tangerang 15310, Indonesia}
  ,altaddress={Department of Physics\footnote{http://www.fisika.ui.ac.id}, University of Indonesia, Kampus UI Depok, Depok 16424, Indonesia}
}

\begin{abstract}
The first cluster-based public computing for Monte Carlo simulation in Indonesia  is introduced. The system has been developed to enable public to perform Monte Carlo simulation on a parallel computer through an integrated and user friendly dynamic web interface. The beta version, so called publicMC@BATAN, has been released and implemented for internal users at the National Nuclear Energy Agency (BATAN). In this paper the concept and architecture of publicMC@BATAN are presented.
\end{abstract}

\maketitle

\section{Introduction}
\label{intro}

Monte Carlo methods are a class of computational algorithms that rely on repeated random sampling to compute their results. The methods are often used when simulating physical and mathematical systems, especially to solve mathematical problems which are unfeasible or impossible to perform analytically or to obtain an exact result with a deterministic algorithm \cite{lanl}.

Monte Carlo simulation methods are especially useful for modeling phenomena with significant uncertainty in inputs and in studying systems with a large number of coupled degrees of freedom, such as fluids, disordered materials, strongly coupled solids, and cellular structures. More broadly, the methods play an important role in the calculation of risk in business. These methods are also widely used in mathematics: a classic use is for the evaluation of definite integrals, particularly multidimensional integrals with complicated boundary conditions. In particular, it is a widely successful method in risk analysis when compared with alternative methods or human intuition.

In physical sciences, Monte Carlo methods are very important in computational physics, physical chemistry and related applied sciences. The methods have diverse applications from complicated quantum chromodynamics calculations in high energy physics to designing heat shields and aerodynamic forms. In statistical physics, particularly Monte Carlo molecular modeling is an alternative for computational molecular dynamics as well as to compute statistical field theories of simple particle and polymer models. In experimental particle physics, these methods are used for designing detectors, understanding their behavior and comparing experimental data to theory, or on vastly large scale of the galaxy modelling. Similar application of Monte Carlo methods are seen in designing the nuclear reactors which requires comprehensive simulations of a huge number of nuclear processes. 

The most popular dedicated software to accomplish such simulations is the Monte Carlo N-Particle Transport Code (MCNP) which was developed and owned by Los Alamos National Laboratory (LANL) \cite{mcnp}. It is used primarily for the simulation of nuclear processes, such as fission, but has the capability to simulate particle interactions involving neutrons, photons, and electrons. Monte Carlo N-Particle eXtended (MCNPX) was also developed at Los Alamos National Laboratory.  It extends the capabilities of MCNP4C3 to nearly all particles, nearly all energies, and to nearly all applications without an additional computational time penalty. MCNPX is fully three-dimensional and time dependent. It utilizes the latest nuclear cross section libraries and uses physics models for particle types and energies where tabular data are not available \cite{mcnpx, futuremcnp}. Both codes can be used to judge whether or not nuclear systems are critical and to determine doses from sources, amongst other things \cite{volken}.

Both MCNP and MCNPX are written in Fortran 90, run on PC Windows, Linux and Unix platforms. More importantly, they are fully parallel and compatible on parallel environments like PVM and MPI. Although MCNP's are available as standalone softwares, more comprehensive and complicated simulations indeed require much more computational power, especially in some cases where the excecuting times are highly considered. The best solution to solve this problem is by paralellizing. 

Unfortunately, under present circumstances in Indonesia most of research groups are not able to afford constructing and maintaining a dedicated cluster computer. The main reasons are the lack of,
\begin{itemize}
 \item Human resources :\\
A real, but even small scale, cluster computer requires a dedicated administrator with high skill on and familiar with parallel environment. The skill is crucial to fine-tune the system according to the nature of computations executed on the system. Otherwise, the cluster would not be able to reach the expected performances.
 \item Financial resources :\\
Although the cost of hardwares is getting more reasonable, constructing a cluster system still requires financial backup which is unaffordable for most research groups. It should be reminded that the total cost also includes the special room with appropriate environment and the daily maintainance.
\end{itemize}
On the other hand due to the lack of human resources, especially the experts having the skill on parallel programming, one can predict easily that a cluster system with a specific characteristic to perform particular jobs would suffer from low utilization which leads to high percentages of idle time.

Therefore, following the same approaches of LIPI Public Cluster (LPC) \cite{lpc, lpc2, icrict}, the authors have developed a ``public'' cluster system dedicated for Monte Carlo simulation, namely publicMC@BATAN \cite{pmcbatan}. The detail of this new infrastructure is presented in the following sections. First, the basic concept of publicMC@BATAN is explained, and followed with the architecture to realize it. The paper is concluded with a summary.

\begin{figure}[t]
        \centering 
	\includegraphics[width=11cm,angle=0,trim=0 0 0 0]{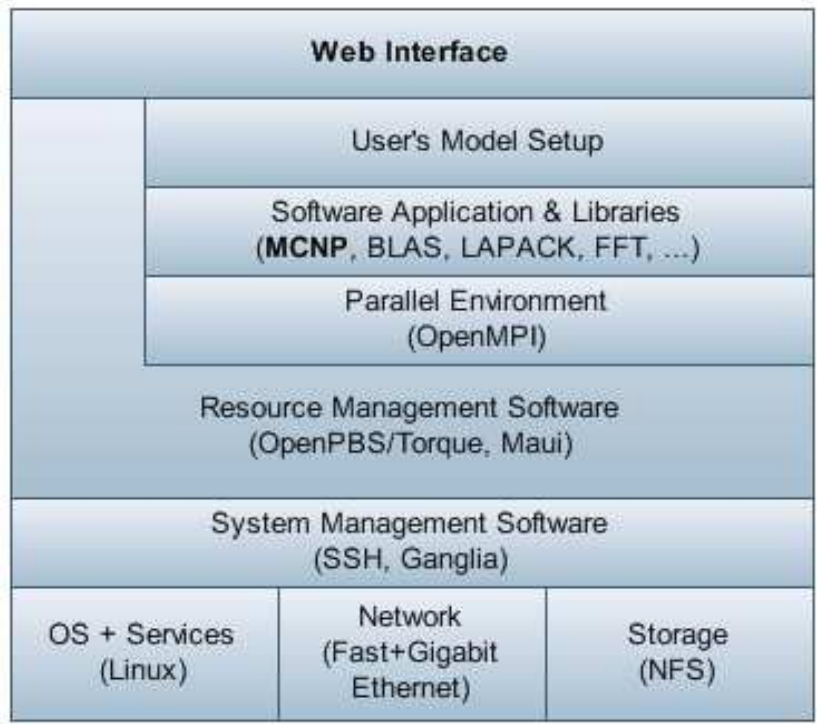}
        \caption{The logical architecture of publicMC@BATAN.}
        \label{fig:pmc}
\end{figure}

\section{Concept}

As already briefly mentioned in the preceeding section, publicMC@BATAN is intended to provide a public infracstructure for Monte Carlo simulation in general. However, the released beta version is yet limited to the Monte Carlo simulation using MCNP packages. This is motivated by the fact that the targeted users in BATAN are using the MCNP, while at time being the system is allowed only for users inside the BATAN network.

Again, publicMC@BATAN has been developed with the following main purposes :
\begin{itemize}
\item Providing a practical but real working place for related users to perform Monte Carlo simulation on parallel environment without having the hardawares at all.
\item Improving the utilization of the existing high performance computing (HPC) facilities. Since in some HPC centers the infrastructures tend to idle away due to limited users.
\end{itemize}
Clearly, these purposes are more relevant for some developing countries with limited HPC resources. However, the issues are, especially the second one is, actually relevant to any other developed regions as well. The limitations are in general caused by tight regulations due to mostly security concerns.

Moreover, the publicMC@BATAN is not designed as a multi purpose public cluster like LPC, but is embedded with particular applications for Monte Carlo simulation, in the present case is MCNP. Therefore, there is no need to divide the nodes strictly in several ``independent'' blocks which is very crucial in a cluster with full-ownership policy as LPC \cite{icts}. In contrast with LPC, publicMC@BATAN deployes the conventional system where all nodes may belong to all active users and the submitted jobs are distributed over all of them using certain a job management system as shown in Fig. \ref{fig:pmc}.

Since all users access the system over the web through HTTP, the issue of security is relevant only at the gateway, or more precisely at the web interface. This makes the whole architecture is much simpler than LPC. This can be accomplished by introducing a registration system, requiring login for registered users and keep the session of each active user in a certain way.

It should be remarked here that according to the above mentioned concepts, publicMC@BATAN has completely different natures with another existing web-based clustering toolkits to enable ''remote access over web`` like OSCAR \cite{oscar}, or some grid portals for interfacing large scale distributed systems over web \cite{ganglia, corba, gird, gridspace}. On the other hand, it does also not belong to the same category as Globus Toolkit (GT) which is intended to ''interface`` large scale grid computing \cite{gt}. We also recognize some partial web-interfaces for clusters developed by several groups as done in \cite{sce, webmpi}.

\section{ Architecture}

As ilustrated in Fig. \ref{fig:pmc}, the gateway plays an important role in publicMC@BATAN. It is not only used as a master node as conventional systems, but it hosts the main web-based interface and security system. Actually, in the system with high load the master node should be separated from the gateway. Comparing with the web interface at LPC \cite{iiwas}, the web interface in publicMC@BATAN has more limited purposes :
\begin{itemize}
\item As a single interface between users and the cluster. This includes integrating all components and making them accessible to remote users over a user-friendly web interface.
\item Providing as high as possible degree of freedom to users, while on the other hand keeping the whole system totally secured. 
\end{itemize}
Fortunately, there is no concern to keep the user and their jobs stay on the assigned blocks as in LPC. Yet, the web interface is the main filter to make sure that only the allowed commands can pass to the master and nodes. This can be done easily by pre-defining the allowed commands relevant to the installed applications.

When a user submits a particular command through the web interface, the system performs several internal processes :
\begin{enumerate}
\item Recognizing whether the command is related to : the distributed computing, i.e. Portable Batch System (PBS) or the operating system (OS). This selection procedure is crucial to determine how to treat and the forwarded target of each command.
\item A submitted command, before paseed to the I/O / master node, is checked if it belongs to the allowed command list or not. Once the command passes the filter, it is immediately forwarded to the resource management system to be further executed.
\end{enumerate}

Since there is no modification on the parallel environment at publicMC@BATAN, one can make use of conventional method to distribute the submitted jobs. The publicMC@BATAN deployes the Terascale Open-source Resource and QUEue manager (TORQUE) and Maui Scheduler for job scheduler \cite{torque, maui}. TORQUE is actually an extension of PBS or definitely OpenPBS \cite{openpbs}. On the other hand, Maui is a "policy engine" to control "when, where, how" the available resources like processor, memory and disk space are allocated to particular jobs in the queue. Maui does not only provide an automated mechanism to optimize the available resources, but also to monitor the performances and analyze the problems occurred during the running periods.

\section{Summary}

A newly developed public infrastructure, so called publicMC@BATAN, for Monte Carlo simulation was briefly introduced. The publicMC@BATAN will be open for public use, although the beta version is yet limited to the internal users at BATAN. Technical and performance tests are still undergoing for MCNP and MCNPX packages. Further, another Monte Carlo packages are going to be installed based on the user's request. 

In the future, publicMC@BATAN can be connected to the global grid using the same protocol and middleware already developed for LPC \cite{iccce, lpcgrid}. This would provide new opportunities for publicMC@BATAN users to collaborate with their global partners on Monte Carlo simulations.

\begin{theacknowledgments}
The work is partially supported by the Riset Kompetitif LIPI in fiscal year $2009$. AAW and NAP thanks the Group for Theoretical and Computational Physics, Research Center for Physics, Indonesian Institute of Sciences for warm hospitality during this work.
\end{theacknowledgments}

\bibliographystyle{aipproc}
\bibliography{icanse2009}

\end{document}